\documentclass[aps,physrev,reprint,groupedaddress]{revtex4-2}
\usepackage{graphicx}
\usepackage{amsmath,amssymb}
\usepackage[T1]{fontenc}
\usepackage{xcolor}

\begin{document}

\title{Poincar\'e sphere representation of spin-driven polarization encoding in two-dimensional perovskite light sources}

\author{Zijian Gan}
\thanks{Equal contributions.}
\affiliation{Department of Chemistry, University of North Carolina at Chapel Hill, Chapel Hill, NC 27599, USA}

\author{Shuyue Feng}
\thanks{Equal contributions.}
\affiliation{Department of Chemistry, University of North Carolina at Chapel Hill, Chapel Hill, NC 27599, USA}

\author{Camryn J. Gloor}
\affiliation{Department of Chemistry, University of North Carolina at Chapel Hill, Chapel Hill, NC 27599, USA}

\author{Wei You}
\affiliation{Department of Chemistry, University of North Carolina at Chapel Hill, Chapel Hill, NC 27599, USA}

\author{Andrew M. Moran}
\email{ammoran@unc.edu}
\affiliation{Department of Chemistry, University of North Carolina at Chapel Hill, Chapel Hill, NC 27599, USA}

\begin{abstract}
Nonlinear optical light sources enable the generation of photons with polarization states that are intrinsically determined by underlying material dynamics, rather than imposed through external modulation. Here, we investigate how four-wave-mixing signal fields emitted by a representative two-dimensional perovskite system can be leveraged for polarization-encoded quantum communication. The experimentally reconstructed signal field is represented by Stokes-vector trajectories on the Poincaré sphere to establish a predictive framework that connects the optical response to practical figures of merit for spin-driven semiconductor light sources. Applied to a model quantum communication scheme in the weak coherent-state regime, this representation quantifies the resolution of the polarization states generated at two pump–probe delays, thereby enabling estimation of the number of detected photons per decoded bit. In addition, a microscopic model connects the signal-field ellipticity to spin relaxation within the exciton fine structure, revealing that interference between resonantly enhanced optical transitions governs the accessible polarization states. The analysis suggests that modest stabilization of the lowest-energy biexciton could substantially improve polarization-encoding performance, offering a route toward materials optimization. More broadly, these results motivate closer integration of nonlinear spectroscopy, semiconductor materials, and quantum information science in the development of novel light sources for quantum communication.
\end{abstract}

\maketitle

\newpage 

Two-dimensional organic–inorganic hybrid perovskites (2D-OIHPs) constitute a versatile materials platform in which optical polarizations can be precisely controlled across the visible spectrum. Their high photoluminescence efficiencies, tunable emission wavelengths, and strong oscillator strengths have enabled applications including light-emitting diodes and optically pumped lasers \cite{yuanPerovskiteEnergyFunnels2016,smithTuningLuminescenceLayered2019,leiEfficientEnergyFunneling2020,alvarado-leañosLasingTwoDimensionalTin2022,linTunableBroadbandMolecular2022}. In addition, the incorporation of chiral molecules within the organic layers between the quantum wells has been shown to support circularly polarized photoluminescence and spin-driven optical effects, with potential relevance for photonic technologies \cite{maChiral2DPerovskites2019,liuBrightCircularlyPolarized2023,liuDirectObservationCircularly2024,chenCircularlyPolarizedLight2019,martinRemoteControlSteering2025,dongChiralityInducedSpinSelectivity2025}. While photoluminescence is generally restricted to singly excited states, both single-exciton and biexciton resonances contribute to nonlinear optical responses \cite{elkinsBiexcitonResonancesReveal2017,thouinStableBiexcitonsTwodimensional2018,bourelleHowExcitonInteractions2020,kochSpectroscopicSignaturesBiexcitons2025}. Under spin-selective circularly polarized excitation, four-wave-mixing signal fields involving biexcitonic states can exhibit high ellipticities, although these effects are generally short-lived in thin 2D-OIHP quantum wells, where spin relaxation occurs on the $\sim$100~fs timescale \cite{toddDetectionRashbaSpin2019,taoDynamicPolaronicScreening2020,chenTuningSpinPolarizedLifetime2021,songRolePolaronicStates2023,romanoCationTuningPolaron2025,soniMechanisticInsightTunable2025}. These material properties were exploited in our recent proof-of-principle quantum communication experiment, in which a material-intrinsic polarization-encoding scheme based on four-wave-mixing signal photons was used to transmit an ASCII message \cite{fengNonlinearOpticalQuantum2026}. Although the emitted fields are coherent states rather than single photons, attenuation to mean photon numbers near unity places them in the weak coherent-state regime employed in established quantum communication protocols, where measurements by an intervening observer can produce detectable disturbances \cite{loDecoyStateQuantum2005,wangBeatingPhotonNumberSplittingAttack2005,curtyInterceptresendAttacksBennettBrassard2005,scaraniSecurityPracticalQuantum2009,bennettQuantumCryptographyPublic2014}.

Herein, we develop an approach in which the polarization states of four-wave-mixing signal fields generated by a 2D-OIHP light source are represented by Stokes vectors on the Poincar\'e sphere. Spin-driven polarization dynamics are mapped onto trajectories across the spherical surface, providing a foundation for metrics that quantify polarization-state distinguishability within communication protocols. Our analysis focuses on the light radiated by the lowest-energy biexciton resonance in the 2D-OIHP material, for which ultrafast spin relaxation drives a broad range of ellipticities \cite{fengNonlinearOpticalSignatures2025,ganMotionalNarrowingSpin2025,fengNonlinearOpticalQuantum2026}. Visualization of the Stokes-vector trajectory associated with this resonance reveals how transformations by optical elements, such as wave plates and polarizers, enable resolution of the encoded states. This framework is applied to a proof-of-principle communication protocol in which bits are encoded in the polarization states of transient-grating signal fields radiated at separate population times. Based on the signal field reconstructed from conventional transient-grating measurements, the resolution of the encoded states is predicted as a function of the number of detected photons, enabling estimation of the number of detected photons per decoded bit. In addition, complementary calculations based on a nonlinear optical response function predict that stabilization of the lowest-energy biexciton resonance improves the resolution of the encoded polarization states for an appropriately chosen measurement basis, providing design principles for future materials.

\begin{figure*}[ht]
    \centering
    \includegraphics[width=0.7\linewidth]{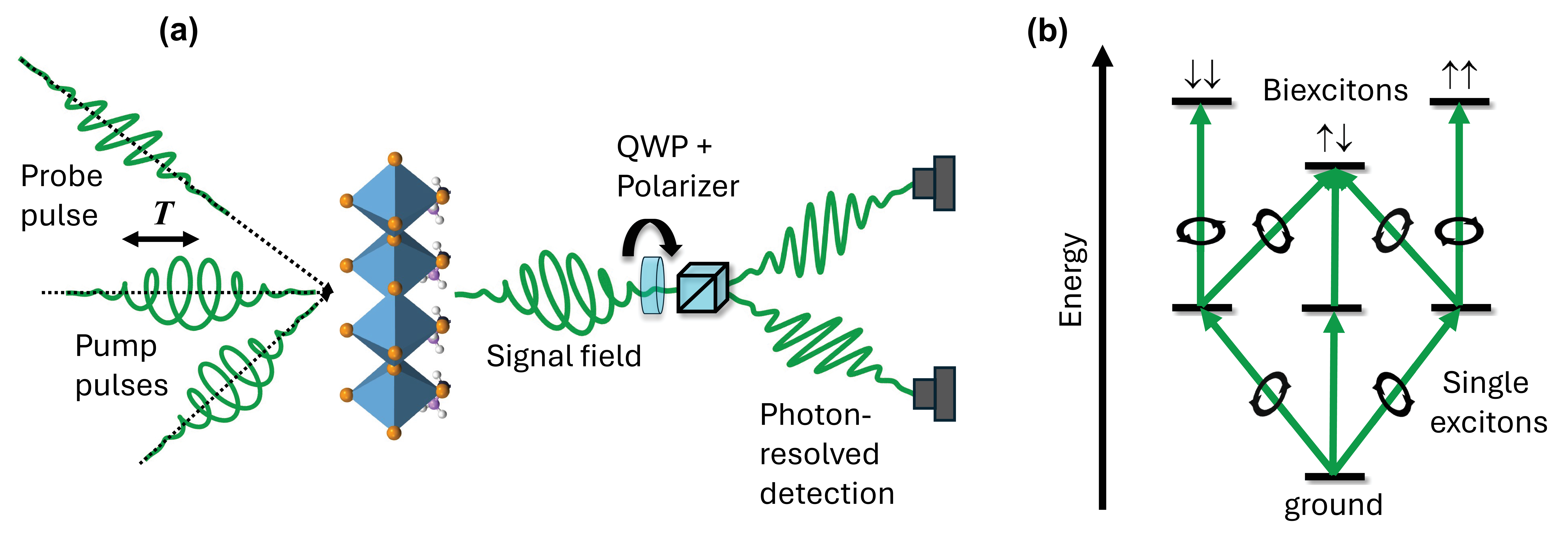}
\caption{Overview of the nonlinear optical quantum communication scheme and the relevant electronic structure. (a) Pump and probe pulses generate a four-wave-mixing signal field whose polarization state is analyzed using a quarter-wave plate (QWP), linear polarizer, and photon-resolved detection. The exciton spin dynamics evolve over the pump--probe delay, $T$, thereby governing the polarization of the emitted signal field. Although not shown schematically, the signal field is spectrally filtered using a 10-nm bandpass before photon-resolved detection to select light radiated within the 535--545~nm spectral region associated with the antiparallel biexciton. (b) Simplified electronic structure showing the effective single-exciton fine-structure manifold together with the parallel- and antiparallel-spin biexciton resonances responsible for the polarization-dependent nonlinear optical response.}
\label{figure1}
\end{figure*}

Figure~\ref{figure1}(a) illustrates the four-wave-mixing communication scheme introduced in Ref.~\cite{fengNonlinearOpticalQuantum2026}. Macroscopic exciton-spin alignment is induced by a pair of circularly polarized pump pulses, while the linearly polarized probe pulse tracks the resulting spin relaxation process through an elliptical-to-linear transformation of the signal field polarization on the $\sim$100~fs timescale \cite{fengNonlinearOpticalSignatures2025}. Attenuation of the signal to a mean photon number near unity enables operation in the weak coherent-state regime relevant to established quantum communication protocols \cite{loDecoyStateQuantum2005,wangBeatingPhotonNumberSplittingAttack2005,curtyInterceptresendAttacksBennettBrassard2005,scaraniSecurityPracticalQuantum2009,bennettQuantumCryptographyPublic2014}. The measurement basis is selected after the sample by randomly toggling the quarter-wave plate angle $\theta_{\mathrm{QWP}}$ between $0^{\circ}$ and $45^{\circ}$, while the pump--probe delay is independently randomized between $T=0$ and $500~\mathrm{fs}$ to generate elliptically and linearly polarized signal photons, respectively. To quantify the polarization contrast under each condition, the signal beam is split by a polarizer and detected with a pair of silicon photomultipliers that resolve the horizontally and vertically polarized photon numbers. Measurements are performed for all four preparation--measurement basis combinations. Following basis reconciliation, the two matching-basis combinations are retained to encode binary 0 and binary 1, whereas the remaining two are discarded. This model communication scheme is loosely inspired by foundational quantum communication protocols, adopting the independent random basis selection of the source (Alice) and the receiver (Bob) as a framework for evaluating the polarization encoding capacity of the 2D-OIHP system \cite{bennettQuantumCryptographyPublic2014}.

The lead--iodide quantum wells employed in the present experiments are separated by phenethylammonium organic layers, forming (PEA)\textsubscript{2}PbI\textsubscript{4}. As indicated in Fig.~\ref{figure1}(b), the system exhibits three single-exciton fine-structure states with either in-plane circularly polarized or out-of-plane linearly polarized transition dipoles. Absorption of circularly polarized light induces a transient population imbalance within the single-exciton manifold, resulting in a macroscopic spin polarization that relaxes through population equilibration within the fine structure. These states are treated as degenerate because the energy splittings are much smaller than the spectroscopic line widths at ambient conditions \cite{sercelExcitonFineStructure2019,posmykQuantificationExcitonFine2022,posmykExcitonFineStructure2024}. The biexciton manifold consists of parallel ($\uparrow\uparrow$ and $\downarrow\downarrow$) and antiparallel ($\uparrow\downarrow$) spin configurations, whose relative energies are determined by the alignment of the single-exciton angular momentum vectors \cite{bourelleHowExcitonInteractions2020,fengNonlinearOpticalSignatures2025}. Based on these single-exciton and biexciton manifolds, the nonlinear optical response can be modeled using three effective resonances rather than by explicitly summing over the individual fine-structure states \cite{fengNonlinearOpticalSignatures2025,ganMotionalNarrowingSpin2025,fengNonlinearOpticalQuantum2026}.

The emitted signal field can be reconstructed as a function of the pump--probe delay through either direct interferometric detection \cite{fengNonlinearOpticalSignatures2025} or iterative fitting of polarization-resolved transient grating signals \cite{fengNonlinearOpticalQuantum2026}. The present analysis employs field amplitudes previously extracted using the latter approach from measurements performed under the same experimental conditions as the model quantum communication experiment. As in the configuration depicted in Fig.~\ref{figure1}, the transient grating measurements used to reconstruct the signal field employed a pair of circularly polarized pump pulses and a vertically polarized probe pulse, while the horizontally and vertically polarized components of the signal were spectrally dispersed onto an array detector. Variation of a quarter-wave plate placed between the sample and detector provided polarization-dependent measurements that constrained a numerical reconstruction of the relative amplitudes and phase of the horizontal and vertical signal-field components. The 20 fs pulses used in these measurements were centered near 520~nm with a spectral width that fully spanned the single-exciton and biexciton resonances.

The electric field radiated by (PEA)\textsubscript{2}PbI\textsubscript{4} is expressed as
\begin{equation}
{{\vec{\mathcal{E} }}_{S}}\left( T,\lambda\right)
=\mathcal{E}_H\left( T,\lambda\right)
{{\hat{e}}_{H}}
+\mathcal{E}_V\left( T,\lambda\right)
{{\hat{e}}_{V}},
\label{eq:Es}
\end{equation}
where $\mathcal{E}_H(T,\lambda)$ and $\mathcal{E}_V(T,\lambda)$ denote the horizontally and vertically polarized complex field amplitudes. The polarization state at each point in the $(T,\lambda)$ space is represented on the Poincar\'e sphere by the normalized Stokes vector $\mathbf{r}=(r_1,r_2,r_3)$, whose components are given by
\begin{equation}
r_1(T,\lambda)
=
\frac{|\mathcal E_H(T,\lambda)|^2-|\mathcal E_V(T,\lambda)|^2}
{|\mathcal E_H(T,\lambda)|^2+|\mathcal E_V(T,\lambda)|^2},
\label{eq:r1}
\end{equation}
\begin{equation}
r_2(T,\lambda)
=
\frac{2\,\mathrm{Re}\!\left[
\mathcal E_H(T,\lambda)
\mathcal E_V^{*}(T,\lambda)
\right]}
{|\mathcal E_H(T,\lambda)|^2+|\mathcal E_V(T,\lambda)|^2},
\label{eq:r2}
\end{equation}
and
\begin{equation}
r_3(T,\lambda)
=
\frac{2\,\mathrm{Im}\!\left[
\mathcal E_H(T,\lambda)
\mathcal E_V^{*}(T,\lambda)
\right]}
{|\mathcal E_H(T,\lambda)|^2+|\mathcal E_V(T,\lambda)|^2}.
\label{eq:r3}
\end{equation}
The $r_1$ component represents the normalized intensity contrast between the horizontal and vertical polarization states, whereas $r_2$ and $r_3$ describe the real and imaginary parts of their mutual coherence. 

The quarter-wave plate located between the sample and detector in Fig.~\ref{figure1}(a) modifies the polarization state by introducing a relative phase shift of $\pi/2$ between orthogonal components. The corresponding transformation of the Stokes vector is given by
\begin{equation}
\mathbf r'
=
\mathbf M_{\rm QWP}(\theta_{\rm QWP})\,
\mathbf r,
\label{eq:mueller_transform}
\end{equation}
where $\theta_{\rm QWP}$ is the angle between the fast axis of the quarter-wave plate and the horizontal polarization axis. The Mueller matrix $\mathbf M_{\rm QWP}$ is given in the Supplementary Material. For the two quarter-wave plate angles employed in this work, the transformed Stokes vector assumes particularly simple forms. For a quarter-wave plate oriented at $\theta_{\rm QWP}=0^\circ$,
\begin{equation}
\mathbf r'
=
\mathbf M_{\rm QWP}(0^\circ)\,
\mathbf r
=
(r_1,r_3,-r_2),
\label{eq:stokes_qwp_0}
\end{equation}
whereas for a quarter-wave plate oriented at $\theta_{\rm QWP}=45^\circ$,
\begin{equation}
\mathbf r'
=
\mathbf M_{\rm QWP}(45^\circ)\,
\mathbf r
=
(r_3,r_2,-r_1).
\label{eq:stokes_qwp_45}
\end{equation}
Together, these two quarter-wave plate settings permit direct measurement of both the $r_1$ and $r_3$ components of the Stokes vector from the horizontal--vertical intensity difference recorded after the linear polarizer.

\begin{figure*}[ht]
\centering
\includegraphics[width=0.85\linewidth]{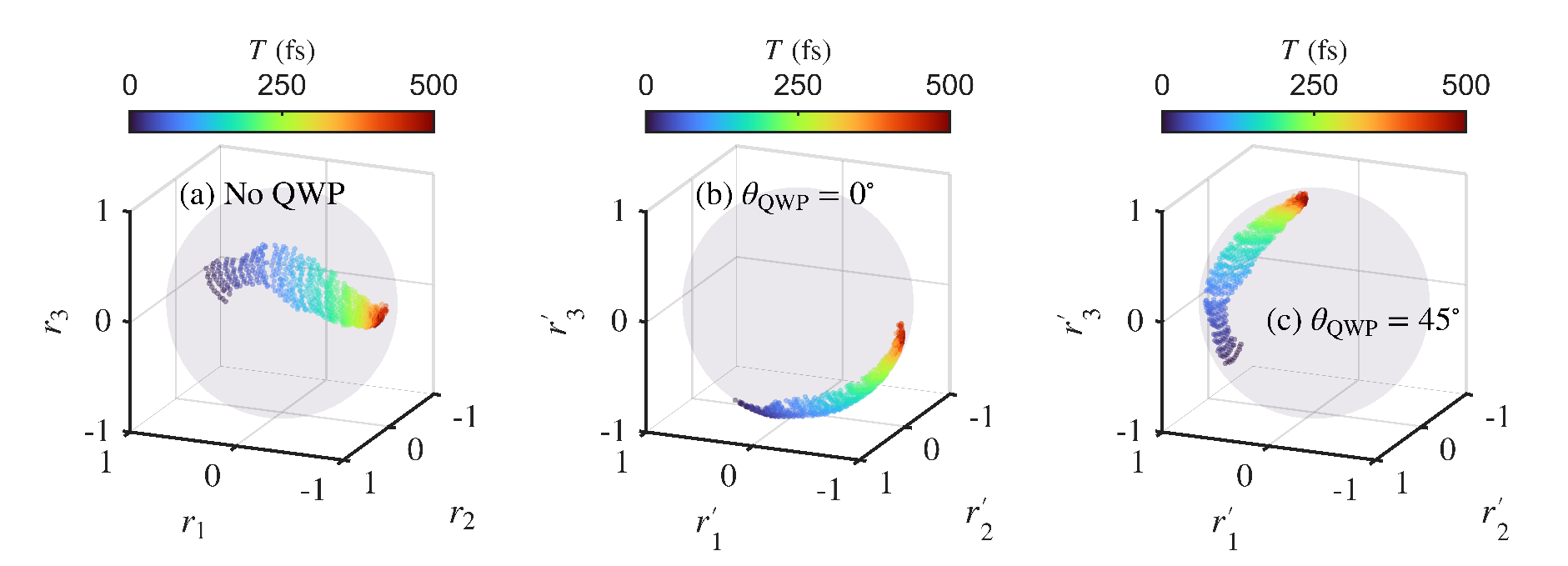}
\caption{Poincar\'e-sphere trajectories computed from the experimentally reconstructed signal field in the 535--545~nm spectral window associated with the antiparallel biexciton. Panels (a)--(c) show the trajectories for no quarter-wave plate, $\theta_{\mathrm{QWP}}=0^\circ$, and $\theta_{\mathrm{QWP}}=45^\circ$, respectively. The colors correspond to the pump--probe delay time, $T$, showing the polarization-state evolution driven by spin relaxation.}
\label{figure2}
\end{figure*}

The experimentally reconstructed signal field is next converted into Stokes-vector trajectories on the surface of the Poincaré sphere, providing a geometric representation of the spin-driven polarization dynamics. To capture the effect of the bandpass filter employed in the quantum communication measurements \cite{fengNonlinearOpticalQuantum2026}, all polarization states within the 535--545~nm wavelength range associated with the antiparallel biexciton are included, thereby broadening the distribution in a direction predominantly transverse to its evolution with the pump--probe delay. In Fig.~\ref{figure2}(a), the trajectory of the directly generated field primarily spreads along the $r_1$ dimension, with spin relaxation driving the polarization state toward $r_1=-1$ (vertical polarization) while progressively reducing the field ellipticity through a decrease in the magnitude of $r_3$. Introducing the quarter-wave plate with orientations of $\theta_{\rm QWP}=0^\circ$ and $45^\circ$ rotates the Stokes-vector trajectories according to Eqs.~\eqref{eq:stokes_qwp_0} and \eqref{eq:stokes_qwp_45}, as illustrated in Figs.~\ref{figure2}(b) and \ref{figure2}(c). Because the communication protocol distinguishes polarization states through their projections onto the horizontal and vertical polarization axes, quarter-wave plate orientations near $\theta_{\rm QWP}=0^\circ$ are expected to improve bit discrimination by maximizing the excursion of the transformed trajectories along the $r_1^\prime$ axis. 

The communication scheme separates the horizontally and vertically polarized (H/V) field components with a polarizer after the quarter-wave plate to distinguish the encoded states. The resulting normalized intensity difference corresponds to the $r_1^\prime$ component of the transformed Stokes vector. At a fixed delay $T$, the polarization state at wavelength $\lambda_k$ is represented by $\mathbf{r}_k^\prime(T)$. Consistent with the reconstructed polarization distributions, each spectral component within the selected wavelength range is assigned equal weight. Therefore, the mean polarization contrast is obtained by averaging these components over the reconstructed polarization ensemble,
\begin{equation}
\overline{P}(T)
=
\frac{1}{N_\lambda}
\sum_{k=1}^{N_\lambda}
r^\prime_{1,k}(T),
\label{eq:Pmean}
\end{equation}
while the corresponding mean-square polarization contrast is given by
\begin{equation}
\left\langle P^2(T)\right\rangle
=
\frac{1}{N_\lambda}
\sum_{k=1}^{N_\lambda}
\left[r^\prime_{1,k}(T)\right]^2.
\label{eq:Pmeansq}
\end{equation}
Broadening of the projected polarization states along the H/V measurement axis is quantified by the variance of the $r_1^\prime$ components,
\begin{equation}
\begin{aligned}
\sigma_P^2(T)
&=
\frac{1}{N_\lambda}
\sum_{k=1}^{N_\lambda}
\left[
r_{1,k}^\prime(T)-\overline{P}(T)
\right]^2 \\
&=
\left\langle P^2(T)\right\rangle-\overline{P}^{\,2}(T).
\end{aligned}
\label{eq:Pvar}
\end{equation}
Together, $\overline{P}(T)$ and $\sigma_P^2(T)$ quantify the delay-dependent evolution of the projected polarization-state distributions.

\begin{figure*}[ht]
    \centering
    \includegraphics[width=0.8\linewidth]{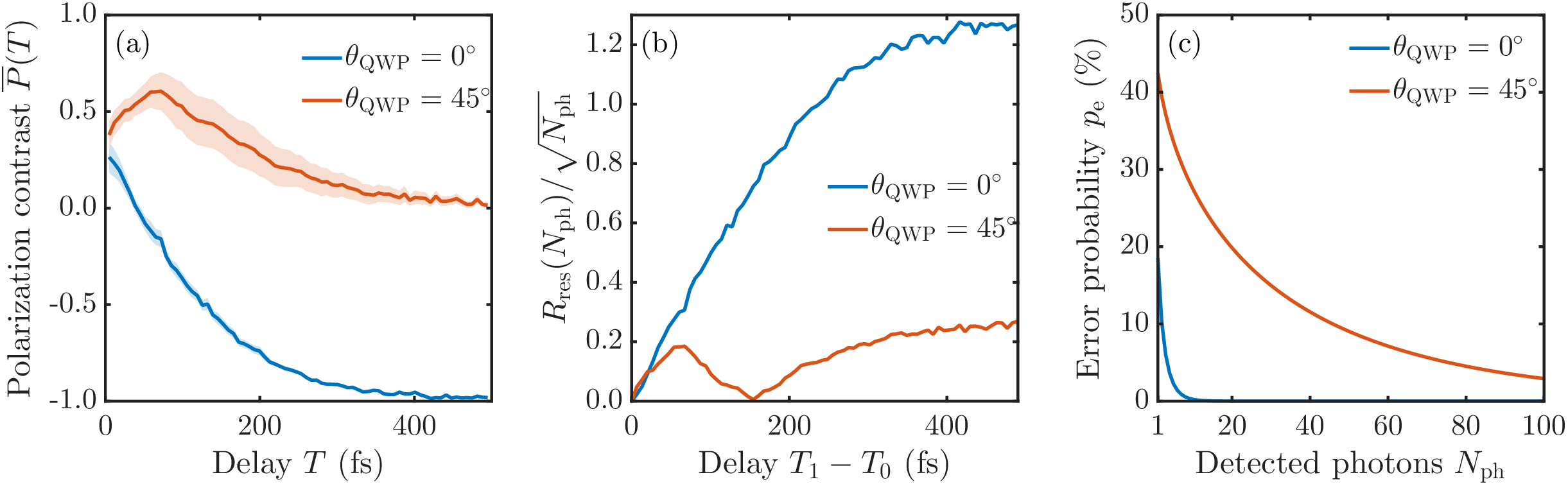}
\caption{Communication performance metrics derived from the experimentally reconstructed signal field over the 535--545~nm spectral window. (a) Mean polarization contrast $\overline{P}(T)$ obtained from the $r_1^\prime$ component of the transformed Stokes vector, with shaded regions indicating the standard deviations $\pm\sigma_P(T)$. (b) Dependence of the resolution parameter $R_{\mathrm{res}}(N_{\mathrm{ph}})/\sqrt{N_{\mathrm{ph}}}$ on the delay separation $T_1-T_0$ for a fixed reference delay of $T_0=0$~fs. (c) Predicted bit-error probability $p_{\mathrm{e}}$ as a function of the number of detected photons for the encoded delay pair $T_0=0$~fs and $T_1=500$~fs.}
\label{figure3}
\end{figure*}

The mean polarization contrasts computed from the experimentally reconstructed signal field are plotted as functions of the population time $T$ for quarter-wave plate orientations of $\theta_{\mathrm{QWP}}=0^\circ$ and $45^\circ$ in Fig.~\ref{figure3}(a). The larger decrease in $\overline{P}(T)$ observed for $\theta_{\mathrm{QWP}}=0^\circ$ directly reflects the greater excursion of the corresponding Stokes-vector trajectory in the $r_1^\prime$ direction. The standard deviation $\sigma_P(T)$ remains negligible for $\theta_{\mathrm{QWP}}=0^\circ$ over the full range of population times because wavelength-dependent polarization broadening occurs mainly in the $r_2^\prime$ dimension. For $\theta_{\mathrm{QWP}}=45^\circ$, the polarization contrast exhibits weaker delay-dependence because the corresponding Stokes-vector trajectory evolves primarily in the $r_3^\prime$ direction, which modulates the relative phase between the H and V field components rather than their measured intensity difference. Additionally, the larger values of $\sigma_P(T)$ observed for $\theta_{\mathrm{QWP}}=45^\circ$ reflect the projection of the polarization-state broadening onto the $r_1^\prime$ axis by the quarter-wave plate transformation.

The distinguishability of two encoded states is determined by the separation of their mean polarization contrasts relative to the statistical variance of the corresponding H/V measurements. The wavelength-averaged H/V photon-counting variance per detected photon is $1-\langle P^2(T)\rangle$, while the wavelength-dependent variation of the polarization contrast contributes $\sigma_P^2(T)$. Because $\sigma_P^2(T)=\langle P^2(T)\rangle-\overline{P}^{\,2}(T)$, these contributions sum to $1-\overline{P}^{\,2}(T)$, the total variance of the wavelength-integrated H/V measurement. For two delays $T_0$ and $T_1$, the resolution in polarization contrast is given by
\begin{equation}
R_{\mathrm{res}}(N_{\mathrm{ph}})
=
\frac{
\sqrt{N_{\mathrm{ph}}}\,
\left|
\overline{P}(T_1)-\overline{P}(T_0)
\right|
}{
\sqrt{
2
-
\overline{P}^{\,2}(T_0)
-
\overline{P}^{\,2}(T_1)
}
},
\label{eq:Rres}
\end{equation}
where $N_{\mathrm{ph}}$ is the number of detected photons. Assuming Gaussian statistics, the probability of incorrectly assigning the received state is
\begin{equation}
p_{\mathrm{e}}
=
\frac{1}{2}
\operatorname{erfc}
\!\left(
\frac{R_{\mathrm{res}}(N_{\mathrm{ph}})}{2}
\right),
\label{eq:Pe}
\end{equation}
which provides a direct connection between the experimentally reconstructed polarization distributions and the predicted communication performance shown in Fig.~\ref{figure3}.

The experimental meaning of \(N_{\rm ph}\) follows directly from the photon-resolved measurements presented in Ref.~\cite{fengNonlinearOpticalQuantum2026}. In those measurements, the signal was attenuated to approximately one detected photon per pulse, and silicon photomultiplier detectors allowed resolution of discrete photoelectron-number levels. Thus, $N_{\rm ph}$ corresponds to the total number of detected photons accumulated over successive laser shots, rather than the number of detection events. For the accumulated polarization contrast obtained from a total of $N_{\rm ph}$ detected photons with fixed polarization contrast $P$, binomial counting statistics give a variance of $(1-P^2)/N_{\rm ph}$, such that the statistical uncertainty depends on the total number of detected photons irrespective of how those photons are distributed among laser shots. The dependence of the polarization contrast and communication resolution on the accumulated number of detected photons was demonstrated experimentally in Ref.~\cite{fengNonlinearOpticalQuantum2026}. Additional details concerning photon-number determination and silicon photomultiplier detector performance are provided in the Supplementary Material.

The polarization-contrast resolution is evaluated by varying the encoding delay $T_1$, with $T_0=0$~fs held fixed, in Fig.~\ref{figure3}(b). Because the achievable resolution scales as $\sqrt{N_{\mathrm{ph}}}$, dividing by this factor yields an intrinsic measure of the polarization-state distinguishability that is independent of the detected photon number. With $\theta_{\mathrm{QWP}}=0^\circ$, $R_{\mathrm{res}}(N_{\mathrm{ph}})$ increases monotonically with $T_1$, consistent with the corresponding polarization-contrast profile shown in Fig.~\ref{figure3}(a). In comparison, the resolution obtained with $\theta_{\mathrm{QWP}}=45^\circ$ exhibits little improvement with increasing delay and is approximately five times smaller at $T_1-T_0=500$~fs. To put these results in the context of the prior communication measurements, all experimental trials conducted with delay times of $T_0=0$~fs and $T_1=500$~fs successfully decoded the 56-bit, 8-character ASCII message after accumulating fewer than 10 detected photons per decoded bit with $\theta_{\mathrm{QWP}}=0^\circ$ \cite{fengNonlinearOpticalQuantum2026}. This performance was typically achieved at a resolution of approximately 3, corresponding to about 6 detected photons per decoded bit and a bit-error probability of $p_{\mathrm{e}}\approx1.7\%$. By contrast, the predicted bit-error probability under the same conditions is approximately 32\% for $\theta_{\mathrm{QWP}}=45^\circ$.

Our model communication experiment provides a common figure of merit, the number of detected photons per decoded bit, for characterizing and comparing the polarization-encoding capacities of different electronic resonances and semiconductor materials in the weak coherent-state regime. The approach is not intended as a complete quantum key distribution (QKD) implementation and is not supported by a security proof. Although the polarization states are represented by reconstructed classical field amplitudes, attenuation to a mean photon number near unity places a finite limit on the polarization information obtainable from each transmitted pulse, distinguishing this regime from the high-intensity limit. A weak coherent pulse contains a Poisson-distributed number of photons with mean $\bar n$, which is distinct from the total detected photon number $N_{\rm ph}$ accumulated by the receiver in decoding a bit. While QKD protocol development is beyond the scope of this work, an intercept--resend calculation can be used to illustrate one distinction between communication in the finite-photon, weak coherent-state regime and the high-intensity limit \cite{curtyInterceptresendAttacksBennettBrassard2005}. At small $\bar n$, the finite number of photons available in each intercepted pulse limits the precision with which an eavesdropper can discriminate the transmitted polarization state, whereas increasing $N_{\rm ph}$ improves the receiver's discrimination of the received states and reduces the bit-error impact of a given disturbance. As shown in the Supplementary Material, the calculation based on our model protocol gives an eavesdropper state-assignment error of approximately 27\% for $\bar n=1$, with the resulting disturbance producing a receiver bit-error probability of approximately 12\% after $N_{\rm ph}=20$ photons have been detected. For the present communication experiment, this analysis illustrates the significance of finite photon number for polarization-encoded quantum communication.

\begin{figure*}[ht]
    \centering
    \includegraphics[width=0.65\linewidth]{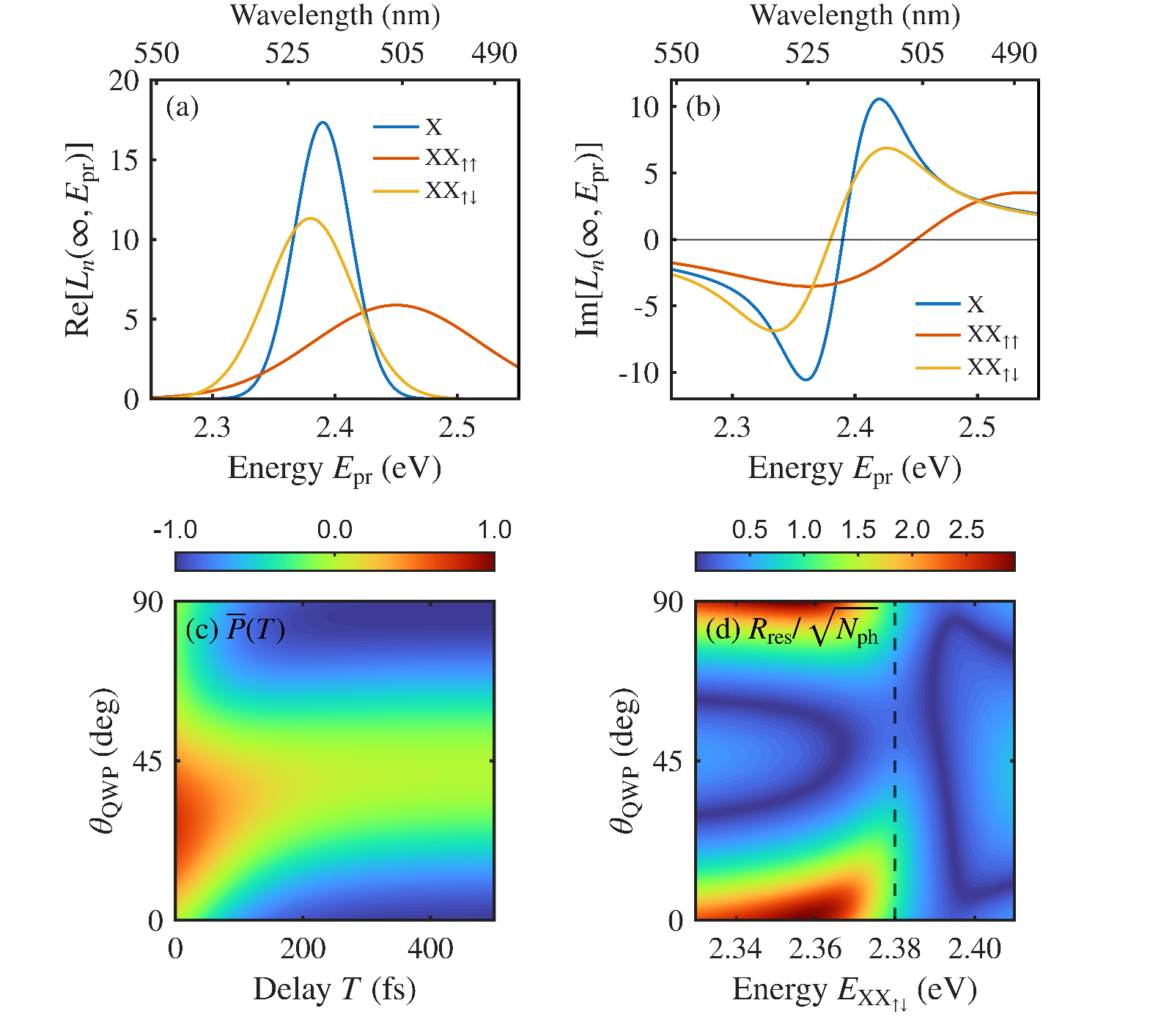}
\caption{Calculated polarization contrast and communication resolution predicted from the nonlinear response function using microscopic parameters. (a) Real and (b) imaginary components of the fully relaxed exciton and biexciton spectral line shapes $L_n(\infty,E_{\rm pr})$. (c) Mean polarization contrast $\overline{P}(T)$ as a function of the pump--probe delay time $T$ and quarter-wave plate orientation $\theta_{\rm QWP}$, averaged over the 535--545~nm probe-energy range. (d) Normalized communication resolution $R_{\mathrm{res}}(N_{\mathrm{ph}})/\sqrt{N_{\rm ph}}$ as a function of the antiparallel biexciton resonance energy $E_{\mathrm{XX}_{\uparrow\downarrow}}$ and quarter-wave plate orientation for the population-time encoding protocol with $T_0=0$~fs and $T_1=500$~fs. The dashed vertical line denotes the experimental value $E_{\mathrm{XX}_{\uparrow\downarrow}}=2.38$~eV.}
\label{figure4}
\end{figure*}

Having established these communication metrics using the experimentally reconstructed signal field, we next employ a theoretical model for the nonlinear response to examine how modifications of the underlying exciton and biexciton electronic structure could alter the polarization-encoding performance. The reconstructed field is used for the quantitative analysis of the measured system in Fig.~3, whereas the model field introduced below is used to explore materials-engineering strategies and generalization to other semiconductor systems.
In prior work, we parameterized the microscopic nonlinear response function of (PEA)\textsubscript{2}PbI\textsubscript{4} in the basis of its exciton fine-structure states using ultrafast spectroscopies \cite{ganElucidatingPhononDephasing2024,fengNonlinearOpticalSignatures2025,ganMotionalNarrowingSpin2025}. However, because the splittings within the fine structure are much smaller than the spectroscopic linewidths at ambient temperatures, the response can be represented by three effective Gaussian line shapes corresponding to the single-exciton ($X$) resonance together with the parallel ($XX_{\uparrow\uparrow}$ and $XX_{\downarrow\downarrow}$) and antiparallel ($XX_{\uparrow\downarrow}$) biexciton resonances \cite{fengNonlinearOpticalQuantum2026}. Therefore, the transient-grating signal field with polarization $j\in\{H,V\}$ is represented as a superposition of these effective resonances,
\begin{equation}
\begin{aligned}
\mathcal E_j(T,E_{\mathrm{pr}})
&=
A_{j,\mathrm{X}}(T)\,L_{\mathrm{X}}(T,E_{\mathrm{pr}}) \\
&\quad
+
A_{j,\mathrm{XX}_{\uparrow\uparrow}}(T)\,L_{\mathrm{XX}_{\uparrow\uparrow}}(T,E_{\mathrm{pr}}) \\
&\quad
+
A_{j,\mathrm{XX}_{\uparrow\downarrow}}(T)\,L_{\mathrm{XX}_{\uparrow\downarrow}}(T,E_{\mathrm{pr}}),
\end{aligned}
\label{eq:Ej}
\end{equation}
where $A_{j,n}(T)$ is the complex amplitude describing the relative phase, transition-dipole coupling, and population-time evolution arising from exciton spin relaxation \cite{fengNonlinearOpticalQuantum2026}. Here, $\mathrm{XX}_{\uparrow\uparrow}$ denotes the parallel-spin biexciton manifold, comprising the $\uparrow\uparrow$ and $\downarrow\downarrow$ configurations, which are degenerate but couple to opposite circular polarizations. These line shape functions and model parameters are summarized in the Supplementary Material.

The real and imaginary parts of the fully relaxed ($T\to\infty$) effective line-shape functions are shown in Figs.~\ref{figure4}(a) and \ref{figure4}(b). The single-exciton and antiparallel biexciton resonances exhibit substantial spectral overlap because of their similar transition energies and linewidths. Because these resonances correspond to singly and doubly excited states, they contribute to the nonlinear response with opposite signs, leading to destructive interference. By contrast, the parallel biexciton resonance is substantially broader and exhibits lower amplitude within the 535--545~nm spectral window under consideration. Analogous to an ultrafast Faraday rotation experiment \cite{baumbergFemtosecondFaradayRotation1994,bourelleOpticalControlExciton2022,sutcliffeFemtosecondMagneticCircular2021}, excitation with circularly polarized pump and vertically polarized probe pulses induces a $\pi/2$ phase shift in the horizontally polarized component of the signal field, thereby exchanging the absorptive and dispersive line shapes in the measured response \cite{fengNonlinearOpticalSignatures2025,fengNonlinearOpticalQuantum2026}. The resulting spin-driven evolution of the signal-field ellipticity reflects interference among the full set of resonances, with the largest polarization changes occurring on the low-energy side of the $L_{XX_{\uparrow\downarrow}}$ line shape near 540~nm. 

The signal field computed from the microscopic response model is converted into the Stokes-vector components using Eqs.~\eqref{eq:r1}--\eqref{eq:r3}, enabling evaluation of the communication performance metrics. Although the response function was independently parameterized \cite{ganElucidatingPhononDephasing2024,fengNonlinearOpticalQuantum2026,ganMotionalNarrowingSpin2025}, it reproduces the essential features of the polarization contrasts computed from the experimentally reconstructed signal field in Fig.~\ref{figure3}(a) while providing insight into their microscopic origin. For illustration, Fig.~\ref{figure4}(c) shows the calculated polarization contrast as a function of the population time $T$ and quarter-wave plate angle $\theta_{\mathrm{QWP}}$. With $\theta_{\mathrm{QWP}}=0^\circ$, the polarization contrast decreases from 0.1 at $T=0$~fs to $-1.0$ at $T=500$~fs. At $\theta_{\mathrm{QWP}}=45^\circ$, the model predicts a decrease in $\overline{P}(T)$ from 0.5 to 0.0 within the same delay range, in good agreement with results obtained using the experimentally reconstructed field. While the model accurately reproduces the polarization contrasts at the encoding delays of $T=0$ and 500~fs, it predicts a monotonic evolution and does not capture the intermediate maximum observed with $\theta_{\mathrm{QWP}}=45^\circ$ in Fig.~\ref{figure3}(a). 

Signal emission in the 540~nm spectral region of the antiparallel biexciton was previously shown to enable quantum communication with approximately 10-fold fewer photons than the higher-energy resonances. Our nonlinear response function attributes this behavior to interference between the single-exciton and antiparallel biexciton resonances, which gives rise to a short-lived burst of red-shifted, horizontally polarized emission near $T=0$~fs \cite{fengNonlinearOpticalQuantum2026}. Having identified the microscopic origin of this interference effect, we next use the model to predict whether stabilizing the antiparallel biexciton can further enhance the communication performance. To explore this possibility, the resolution parameter $R_{\mathrm{res}}(N_{\mathrm{ph}})$ shown in Fig.~\ref{figure4}(d) is computed using Eq.~\eqref{eq:Rres} while varying $E_{\mathrm{XX}_{\uparrow\downarrow}}$, with all other model parameters held fixed. The calculations suggest that the communication performance can be substantially improved by reducing $E_{\mathrm{XX}_{\uparrow\downarrow}}$ by only 0.01--0.05~eV. The predicted enhancement is largest near $\theta_{\mathrm{QWP}}=0^\circ$, whereas stabilizing the antiparallel biexciton reduces $R_{\mathrm{res}}(N_{\mathrm{ph}})$ at $\theta_{\mathrm{QWP}}=45^\circ$. This dependence on the quarter-wave plate angle reflects the orientations of the Stokes-vector trajectories on the Poincar\'e sphere depicted in Fig.~\ref{figure2}, as $\theta_{\mathrm{QWP}}=0^\circ$ produces near-optimal alignment along the $r_1^\prime$ axis. 

In summary, the present work extends the previous experimental demonstration of nonlinear optical quantum communication using a 2D-OIHP light source by establishing a general predictive framework for analyzing and optimizing spin-driven polarization encoding. The four-wave-mixing signal field is translated into a Stokes-vector trajectory on the surface of the Poincar\'e sphere, directly revealing the distribution of polarization states and their transformations by optical elements. This representation connects the signal field to experimentally accessible communication metrics, allowing the polarization-encoding performance of different electronic resonances and candidate materials to be evaluated with common figures of merit in the weak coherent-state regime. In the present semiconductor light source, resonant nonlinear optical interactions produce wavelength-dependent field amplitudes and phases, thereby generating a distribution of polarization states across the emitted optical spectrum. The width of the polarization-state distribution reveals wavelength-dependent polarization variations that are not captured by the mean polarization contrast alone. This behavior differs from conventional polarization encoding imposed externally on an attenuated laser field using an electro-optic modulator, for which the polarization can be made approximately uniform across the optical bandwidth. Furthermore, the nonlinear response model developed for the 2D-OIHP system suggests that modest stabilization of the antiparallel biexciton energy could substantially enhance polarization-encoding performance for an appropriately chosen measurement basis, providing a potential strategy for materials optimization.

\section*{Supplementary Material}

We present an analysis of the photon-resolved detection process, the Mueller-matrix representation of the quarter-wave plate transformation, an intercept--resend disturbance model for the weak coherent-state regime, and a microscopic model for computing the Stokes-vector components of transient-grating signal fields.

\begin{acknowledgments}
This work is supported by the National Science Foundation under Grant Nos.~CHE-2547606 (S.F., Z.G., and A.M.) and CHE-2154791 (C.G. and W.Y.). C.G. acknowledges support from the U.S. National Science Foundation Graduate Research Fellowship Program under Grant No. DGE-2439854. This work was performed in part at the Chapel Hill Analytical and Nanofabrication Laboratory (CHANL), a member of the North Carolina Research Triangle Nanotechnology Network (RTNN), which is supported by the National Science Foundation under Grant No.~ECCS-2025064, as part of the National Nanotechnology Coordinated Infrastructure (NNCI).
\end{acknowledgments}

\section*{Conflicts of Interest }
The authors have no conflicts to disclose.

\section*{Data Availability}
The data that support the findings of this study are available from the corresponding author upon reasonable request. 

\bibliographystyle{aipnum4-2}
\bibliography{ZoteroLibraryAbbrev}

@article{alvarado-leañosLasingTwoDimensionalTin2022,
  title = {Lasing in {{Two-Dimensional Tin Perovskites}}},
  author = {{Alvarado-Lea{\~n}os}, Ada Lil{\'i} and Cortecchia, Daniele and Saggau, Christian Niclaas and Martani, Samuele and Folpini, Giulia and Feltri, Elena and Albaqami, Munirah D. and Ma, Libo and Petrozza, Annamaria},
  year = 2022,
  month = dec,
  journal = {ACS Nano},
  volume = {16},
  number = {12},
  pages = {20671--20679},
  issn = {1936-0851},
  doi = {10.1021/acsnano.2c07705}
}

@article{baumbergFemtosecondFaradayRotation1994,
  title = {Femtosecond {{Faraday}} Rotation in Spin-engineered Heterostructures},
  author = {Baumberg, J. J. and Awschalom, D. D. and Samarth, N.},
  year = 1994,
  journal = {J. Appl. Phys.},
  volume = {75},
  number = {10},
  pages = {6199--6204},
  issn = {0021-8979},
  doi = {10.1063/1.355455}
}

@article{bennettQuantumCryptographyPublic2014,
  title = {Quantum Cryptography: {{Public}} Key Distribution and Coin Tossing},
  author = {Bennett, Charles H. and Brassard, Gilles},
  year = {2014},
  journal = {Theor. Comput. Sci.},
  volume = {560},
  pages = {7--11},
  issn = {0304-3975},
  doi = {10.1016/j.tcs.2014.05.025}
}

@article{bourelleHowExcitonInteractions2020,
  title = {How {{Exciton Interactions Control Spin-Depolarization}} in {{Layered Hybrid Perovskites}}},
  author = {Bourelle, Sean A. and Shivanna, Ravichandran and Camargo, Franco V. A. and Ghosh, Soumen and Gillett, Alexander J. and Senanayak, Satyaprasad P. and Feldmann, Sascha and Eyre, Lissa and Ashoka, Arjun and {van de Goor}, Tim W. J. and Abolins, Haralds and Winkler, Thomas and Cerullo, Giulio and Friend, Richard H. and Deschler, Felix},
  year = 2020,
  month = aug,
  journal = {Nano Lett.},
  volume = {20},
  number = {8},
  pages = {5678--5685},
  issn = {1530-6984},
  doi = {10.1021/acs.nanolett.0c00867}
}

@article{bourelleOpticalControlExciton2022,
  title = {Optical Control of Exciton Spin Dynamics in Layered Metal Halide Perovskites via Polaronic State Formation},
  author = {Bourelle, Sean A. and Camargo, Franco V. A. and Ghosh, Soumen and Neumann, Timo and {van de Goor}, Tim W. J. and Shivanna, Ravichandran and Winkler, Thomas and Cerullo, Giulio and Deschler, Felix},
  year = 2022,
  month = jun,
  journal = {Nat. Commun.},
  volume = {13},
  number = {1},
  pages = {3320},
  issn = {2041-1723},
  doi = {10.1038/s41467-022-30953-w}
}

@article{chenCircularlyPolarizedLight2019,
  title = {Circularly Polarized Light Detection Using Chiral Hybrid Perovskite},
  author = {Chen, Chao and Gao, Liang and Gao, Wanru and Ge, Cong and Du, Xinyuan and Li, Zha and Yang, Ying and Niu, Guangda and Tang, Jiang},
  year = 2019,
  month = apr,
  journal = {Nat. Commun.},
  volume = {10},
  number = {1},
  pages = {1927},
  issn = {2041-1723},
  doi = {10.1038/s41467-019-09942-z}
}

@article{chenTuningSpinPolarizedLifetime2021,
  title = {Tuning {{Spin-Polarized Lifetime}} in {{Two-Dimensional Metal}}--{{Halide Perovskite}} through {{Exciton Binding Energy}}},
  author = {Chen, Xihan and Lu, Haipeng and Wang, Kang and Zhai, Yaxin and Lunin, Vladimir and Sercel, Peter C. and Beard, Matthew C.},
  year = 2021,
  month = nov,
  journal = {J. Am. Chem. Soc.},
  volume = {143},
  number = {46},
  pages = {19438--19445},
  issn = {0002-7863},
  doi = {10.1021/jacs.1c08514}
}

@article{curtyInterceptresendAttacksBennettBrassard2005,
  title = {Intercept-Resend Attacks in the {{Bennett-Brassard}} 1984 Quantum-Key-Distribution Protocol with Weak Coherent Pulses},
  author = {Curty, Marcos and L{\"u}tkenhaus, Norbert},
  year = 2005,
  month = jun,
  journal = {Phys. Rev. A},
  volume = {71},
  number = {6},
  pages = {062301},
  publisher = {American Physical Society},
  doi = {10.1103/PhysRevA.71.062301},
  urldate = {2026-09-24}
}

@article{dongChiralityInducedSpinSelectivity2025,
  title = {Chirality-{{Induced Spin Selectivity}} in {{Hybrid Organic-Inorganic Perovskite Semiconductors}}},
  author = {Dong, Yifan and Hautzinger, Matthew P. and Haque, Md Azimul and Beard, Matthew C.},
  year = 2025,
  month = apr,
  journal = {Annu. Rev. Phys. Chem.},
  volume = {76},
  number = {Volume 76, 2025},
  pages = {519--537},
  publisher = {Annual Reviews},
  issn = {0066-426X, 1545-1593},
  doi = {10.1146/annurev-physchem-082423-032933},
  urldate = {2025-11-26},
  langid = {english}
}

@article{elkinsBiexcitonResonancesReveal2017,
  title = {Biexciton {{Resonances Reveal Exciton Localization}} in {{Stacked Perovskite Quantum Wells}}},
  author = {Elkins, Madeline H. and Pensack, Ryan and Proppe, Andrew H. and Voznyy, Oleksandr and Quan, Li Na and Kelley, Shana O. and Sargent, Edward H. and Scholes, Gregory D.},
  year = 2017,
  month = aug,
  journal = {J. Phys. Chem. Lett.},
  volume = {8},
  number = {16},
  pages = {3895--3901},
  publisher = {American Chemical Society},
  doi = {10.1021/acs.jpclett.7b01621},
  urldate = {2025-12-20}
}

@article{fengNonlinearOpticalSignatures2025,
  title = {Nonlinear Optical Signatures of Spin Relaxation in {{2D}} Perovskites},
  author = {Feng, Shuyue and Badalis, Christopher J. and Gloor, Camryn J. and Zhong, Xiaowei and Gan, Zijian and You, Wei and Moran, Andrew M.},
  year = 2025,
  journal = {J. Chem. Phys.},
  volume = {162},
  number = {13},
  pages = {134202},
  issn = {0021-9606},
  doi = {10.1063/5.0255426}
}

@article{fengNonlinearOpticalQuantum2026,
  title = {Nonlinear Optical Quantum Communication with a Two-Dimensional Perovskite Light Source},
  author = {Feng, Shuyue and Gan, Zijian and Gloor, Camryn J. and You, Wei and Moran, Andrew M.},
  year = 2026,
  month = feb,
  journal = {J. Chem. Phys.},
  volume = {164},
  number = {7},
  pages = {074203},
  issn = {0021-9606},
  doi = {10.1063/5.0314600},
  urldate = {2026-04-29}
}

@article{ganMotionalNarrowingSpin2025,
  title = {Motional Narrowing of Spin Relaxation in {{2D}} Perovskites by Correlated Exciton Fluctuations},
  author = {Gan, Zijian and Feng, Shuyue and Gloor, Camryn J. and Zhong, Xiaowei and You, Wei and Moran, Andrew M.},
  year = 2025,
  month = oct,
  journal = {J. Chem. Phys.},
  volume = {163},
  number = {16},
  pages = {164203},
  issn = {0021-9606},
  doi = {10.1063/5.0293857},
  urldate = {2025-11-16}
}

@article{ganElucidatingPhononDephasing2024,
  title = {Elucidating Phonon Dephasing Mechanisms in Layered Perovskites with Coherent {{Raman}} Spectroscopies},
  author = {Gan, Zijian and Gloor, Camryn J. and Yan, Liang and Zhong, Xiaowei and You, Wei and Moran, Andrew M.},
  year = 2024,
  journal = {J. Chem. Phys.},
  volume = {161},
  number = {7},
  pages = {074202},
  issn = {0021-9606},
  doi = {10.1063/5.0216472}
}

@article{kochSpectroscopicSignaturesBiexcitons2025,
  title = {Spectroscopic Signatures of Biexcitons: {{A}} Case Study in {{Ruddlesden}}--{{Popper}} Lead-Halides},
  shorttitle = {Spectroscopic Signatures of Biexcitons},
  author = {Koch, Katherine A. and {Rojas-Gatjens}, Esteban and {G{\'o}mez-Dominguez}, Mart{\'i}n and {Correa-Baena}, Juan-Pablo and {Silva-Acu{\~n}a}, Carlos and Srimath Kandada, Ajay Ram},
  year = 2025,
  month = jul,
  journal = {J. Chem. Phys.},
  volume = {163},
  number = {3},
  pages = {034202},
  issn = {0021-9606},
  doi = {10.1063/5.0271075},
  urldate = {2026-05-24}
}

@article{leiEfficientEnergyFunneling2020,
  title = {Efficient {{Energy Funneling}} in {{Quasi-2D Perovskites}}: {{From Light Emission}} to {{Lasing}}},
  author = {Lei, Lei and Seyitliyev, Dovletgeldi and Stuard, Samuel and Mendes, Juliana and Dong, Qi and Fu, Xiangyu and Chen, Yi-An and He, Siliang and Yi, Xueping and Zhu, Liping and Chang, Chih-Hao and Ade, Harald and Gundogdu, Kenan and So, Franky},
  year = 2020,
  month = apr,
  journal = {Adv. Mater.},
  volume = {32},
  number = {16},
  pages = {1906571},
  issn = {0935-9648},
  doi = {10.1002/adma.201906571}
}

@article{linTunableBroadbandMolecular2022,
  title = {Tunable {{Broadband Molecular Emission}} in {{Mixed-Organic-Cation Two-Dimensional Hybrid Perovskites}}},
  author = {Lin, YunHui L. and Johnson, Justin C.},
  year = 2022,
  month = nov,
  journal = {ACS Appl. Opt. Mater.},
  volume = {1},
  number = {1},
  pages = {3--9},
  issn = {2771-9855},
  doi = {10.1021/acsaom.2c00106},
  urldate = {2026-07-31}
}

@article{liuBrightCircularlyPolarized2023,
  title = {Bright Circularly Polarized Photoluminescence in Chiral Layered Hybrid Lead-Halide Perovskites},
  author = {Liu, Shangpu and Kepenekian, Mika{\"e}l and Bodnar, Stanislav and Feldmann, Sascha and Heindl, Markus W. and Fehn, Natalie and Zerhoch, Jonathan and Shcherbakov, Andrii and P{\"o}thig, Alexander and Li, Yang and Paetzold, Ulrich W. and Kartouzian, Aras and Sharp, Ian D. and Katan, Claudine and Even, Jacky and Deschler, Felix},
  year = 2023,
  journal = {Sci. Adv.},
  volume = {9},
  number = {35},
  pages = {eadh5083},
  doi = {doi:10.1126/sciadv.adh5083}
}

@article{liuDirectObservationCircularly2024,
  title = {Direct {{Observation}} of {{Circularly Polarized Nonlinear Optical Activities}} in {{Chiral Hybrid Lead Halides}}},
  author = {Liu, Sunhao and Wang, Xiaoming and Dou, Yixuan and Wang, Qian and Kim, Jiyoon and Slebodnick, Carla and Yan, Yanfa and Quan, Lina},
  year = 2024,
  month = may,
  journal = {J. Am. Chem. Soc.},
  volume = {146},
  number = {17},
  pages = {11835--11844},
  issn = {0002-7863},
  doi = {10.1021/jacs.4c00619}
}

@article{loDecoyStateQuantum2005,
  title = {Decoy {{State Quantum Key Distribution}}},
  author = {Lo, Hoi-Kwong and Ma, Xiongfeng and Chen, Kai},
  year = {2005},
  journal = {Phys. Rev. Lett.},
  volume = {94},
  number = {23},
  pages = {230504},
  doi = {10.1103/PhysRevLett.94.230504}
}

@article{maChiral2DPerovskites2019,
  title = {Chiral {{2D Perovskites}} with a {{High Degree}} of {{Circularly Polarized Photoluminescence}}},
  author = {Ma, Jiaqi and Fang, Chen and Chen, Chao and Jin, Long and Wang, Jiaqi and Wang, Shuai and Tang, Jiang and Li, Dehui},
  year = 2019,
  month = mar,
  journal = {ACS Nano},
  volume = {13},
  number = {3},
  pages = {3659--3665},
  issn = {1936-0851},
  doi = {10.1021/acsnano.9b00302}
}

@article{martinRemoteControlSteering2025,
  title = {Remote Control: {{Steering}} Chiroptical Activity in Perovskites via Chiral Additives},
  shorttitle = {Remote Control},
  author = {Martin, Perry W. and Bischak, Connor G.},
  year = 2025,
  month = aug,
  journal = {Matter},
  volume = {8},
  number = {8},
 pages = {102297},
  publisher = {Elsevier},
  issn = {2590-2393, 2590-2385},
  doi = {10.1016/j.matt.2025.102297},
  urldate = {2025-11-26},
  langid = {english}
}

@article{posmykExcitonFineStructure2024,
  title = {Exciton {{Fine Structure}} in {{2D Perovskites}}: {{The Out-of-Plane Excitonic State}}},
  author = {Posmyk, Katarzyna and Dyksik, Mateusz and Surrente, Alessandro and Maude, Duncan K. and Zawadzka, Natalia and Babi{\'n}ski, Adam and Molas, Maciej R. and Paritmongkol, Watcharaphol and M{\k a}czka, Miros{\l}aw and Tisdale, William A. and Plochocka, Paulina and Baranowski, Micha{\l}},
  year = 2024,
  month = mar,
  journal = {Adv. Opt. Mater.},
  volume = {12},
  number = {8},
  pages = {2300877},
  issn = {2195-1071},
  doi = {10.1002/adom.202300877}
}

@article{posmykQuantificationExcitonFine2022,
  title = {Quantification of {{Exciton Fine Structure Splitting}} in a {{Two-Dimensional Perovskite Compound}}},
  author = {Posmyk, Katarzyna and Zawadzka, Natalia and Dyksik, Mateusz and Surrente, Alessandro and Maude, Duncan K. and Kazimierczuk, Tomasz and Babi{\'n}ski, Adam and Molas, Maciej R. and Paritmongkol, Watcharaphol and M{\k a}czka, Miros{\l}aw and Tisdale, William A. and P{\l}ochocka, Paulina and Baranowski, Micha{\l}},
  year = 2022,
  month = may,
  journal = {J. Phys. Chem. Lett.},
  volume = {13},
  number = {20},
  pages = {4463--4469},
  doi = {10.1021/acs.jpclett.2c00942}
}

@article{romanoCationTuningPolaron2025,
  title = {Cation {{Tuning}} of {{Polaron Barriers}} in {{Layered Perovskites}} for {{Optical Spin Lifetime Control}}},
  author = {Romano, Valentino and H{\"o}rmann, Martin and Stadlbauer, Anna and Mosconi, Edoardo and Gregori, Luca and De Angelis, Filippo and Deschler, Felix and Cerullo, Giulio and Camargo, Franco V. A.},
  year = 2025,
  month = sep,
  journal = {ACS Energy Lett.},
  volume = {10},
  number = {9},
  pages = {4636--4643},
  publisher = {American Chemical Society},
  doi = {10.1021/acsenergylett.5c01236},
  urldate = {2025-11-26}
}

@article{scaraniSecurityPracticalQuantum2009,
  title = {The Security of Practical Quantum Key Distribution},
  author = {Scarani, Valerio and {Bechmann-Pasquinucci}, Helle and Cerf, Nicolas J. and Du{\v s}ek, Miloslav and L{\"u}tkenhaus, Norbert and Peev, Momtchil},
  year = 2009,
  month = sep,
  journal = {Rev. Mod. Phys.},
  volume = {81},
  number = {3},
  pages = {1301--1350},
  publisher = {American Physical Society},
  doi = {10.1103/RevModPhys.81.1301},
  urldate = {2025-12-26}
}

@article{sercelExcitonFineStructure2019,
  title = {Exciton {{Fine Structure}} in {{Perovskite Nanocrystals}}},
  author = {Sercel, Peter C. and Lyons, John L. and Wickramaratne, Darshana and Vaxenburg, Roman and Bernstein, Noam and Efros, Alexander L.},
  year = 2019,
  month = jun,
  journal = {Nano Lett.},
  volume = {19},
  number = {6},
  pages = {4068--4077},
  issn = {1530-6984},
  doi = {10.1021/acs.nanolett.9b01467}
}

@article{smithTuningLuminescenceLayered2019,
  title = {Tuning the {{Luminescence}} of {{Layered Halide Perovskites}}},
  author = {Smith, Matthew D. and Connor, Bridget A. and Karunadasa, Hemamala I.},
  year = 2019,
  month = mar,
  journal = {Chem. Rev.},
  volume = {119},
  number = {5},
  pages = {3104--3139},
  issn = {0009-2665},
  doi = {10.1021/acs.chemrev.8b00477}
}

@article{songRolePolaronicStates2023,
  title = {The {{Role}} of {{Polaronic States}} on the {{Spin Dynamics}} in {{Solution-Processed Two-Dimensional Layered Perovskite}} with {{Different Layer Thickness}}},
  author = {Song, Mu-Sen and Wang, Hai and Hu, Zi-Fan and Zhang, Yu-Peng and Liu, Tian-Yu and Wang, Hai-Yu},
  year = 2023,
  journal = {Adv. Sci.},
  volume = {10},
  number = {26},
  pages = {2302554},
  issn = {2198-3844},
  doi = {10.1002/advs.202302554}
}

@article{soniMechanisticInsightTunable2025,
  title = {Mechanistic {{Insight}} into {{Tunable Spin Relaxation}} in {{Two-Dimensional Type-II Ligand-Perovskite Heterostructures}}},
  author = {Soni, Ashish and Yang, Cheng and Yang, Yu-Ting and Lin, Chenjian and Dou, Letian and Wang, Lili},
  year = 2025,
  month = nov,
  journal = {J. Am. Chem. Soc.},
  volume = {147},
  number = {45},
  pages = {41845--41854},
  publisher = {American Chemical Society},
  issn = {0002-7863},
  doi = {10.1021/jacs.5c14647},
  urldate = {2025-12-20}
}

@article{sutcliffeFemtosecondMagneticCircular2021,
  title = {A Femtosecond Magnetic Circular Dichroism Spectrometer},
  author = {Sutcliffe, Jake and Johansson, J. Olof},
  year = 2021,
  journal = {Rev. Sci. Instrum.},
  volume = {92},
  number = {11},
  pages = {113001},
  issn = {0034-6748},
  doi = {10.1063/5.0064460}
}

@article{taoDynamicPolaronicScreening2020,
  title = {Dynamic Polaronic Screening for Anomalous Exciton Spin Relaxation in Two-Dimensional Lead Halide Perovskites},
  author = {Tao, Weijian and Zhou, Qiaohui and Zhu, Haiming},
  year = 2020,
  journal = {Sci. Adv.},
  volume = {6},
  number = {47},
  pages = {eabb7132},
  doi = {doi:10.1126/sciadv.abb7132}
}

@article{thouinStableBiexcitonsTwodimensional2018,
  title = {Stable Biexcitons in Two-Dimensional Metal-Halide Perovskites with Strong Dynamic Lattice Disorder},
  author = {Thouin, F{\'e}lix and Neutzner, Stefanie and Cortecchia, Daniele and Dragomir, Vlad Alexandru and Soci, Cesare and Salim, Teddy and Lam, Yeng Ming and Leonelli, Richard and Petrozza, Annamaria and Kandada, Ajay Ram Srimath and Silva, Carlos},
  year = {2018},
  journal = {Phys. Rev. Mater.},
  volume = {2},
  number = {3},
  pages = {034001},
  doi = {10.1103/PhysRevMaterials.2.034001}
}

@article{toddDetectionRashbaSpin2019,
  title = {Detection of {{Rashba}} Spin Splitting in {{2D}} Organic-Inorganic Perovskite via Precessional Carrier Spin Relaxation},
  author = {Todd, Seth B. and Riley, Drew B. and {Binai-Motlagh}, Ali and Clegg, Charlotte and Ramachandran, Ajan and March, Samuel A. and Hoffman, Justin M. and Hill, Ian G. and Stoumpos, Constantinos C. and Kanatzidis, Mercouri G. and Yu, Zhi-Gang and Hall, Kimberley C.},
  year = 2019,
  journal = {APL Materials},
  volume = {7},
  number = {8},
  pages = {081116},
  issn = {2166-532X},
  doi = {10.1063/1.5099352}
}

@article{wangBeatingPhotonNumberSplittingAttack2005,
  title = {Beating the {{Photon-Number-Splitting Attack}} in {{Practical Quantum Cryptography}}},
  author = {Wang, Xiang-Bin},
  year = {2005},
  journal = {Phys. Rev. Lett.},
  volume = {94},
  number = {23},
  pages = {230503},
  doi = {10.1103/PhysRevLett.94.230503}
}

@article{yuanPerovskiteEnergyFunnels2016,
  title = {Perovskite Energy Funnels for Efficient Light-Emitting Diodes},
  author = {Yuan, Mingjian and Quan, Li Na and Comin, Riccardo and Walters, Grant and Sabatini, Randy and Voznyy, Oleksandr and Hoogland, Sjoerd and Zhao, Yongbiao and Beauregard, Eric M. and Kanjanaboos, Pongsakorn and Lu, Zhenghong and Kim, Dong Ha and Sargent, Edward H.},
  year = 2016,
  month = oct,
  journal = {Nat. Nanotechnol.},
  volume = {11},
  number = {10},
  pages = {872--877},
  issn = {1748-3395},
  doi = {10.1038/nnano.2016.110}
}

\end{document}